\RequirePackage[separate-uncertainty = true,multi-part-units=single]{siunitx}
\documentclass[doublecol]{epl2} 

\usepackage[utf8]{inputenc}

\usepackage{amssymb}
\usepackage{mathtools}

\usepackage{hyperref}

\title{Time-dependent entropy of a cooling Bose gas}

\author{Georg Wolschin}
\shortauthor{G. Wolschin}

\institute{                    
  Institut f{\"ur} Theoretische
Physik
der Universit{\"a}t Heidelberg, Philosophenweg 12-16, D-69120 Heidelberg, Germany, EU\\}

\abstract{Exact analytic solutions of a nonlinear boson diffusion equation with suitable initial conditions that account for
evaporative cooling of ultracold atoms, plus boundary conditions at the singularity $\epsilon=\mu<0$ are presented, and used to calculate the time-dependent
entropy of a cold quantum gas. 
} 

\begin{document}

\maketitle

\section{Introduction}
Thermalization processes are of great interest in physics across the energy scale from pico- to teraelectronvolts. Basic examples for bosonic systems are
the fast approach of gluons to local equilibrium in the initial stages of a relativistic heavy-ion collision at the Large Hadron Collider, or the thermal equilibration of cold bosonic atoms
such as $^{23}$Na or $^{87}$Rb  in the course of evaporative cooling \cite{BookPitaevskii}. Due to the possibility of condensate formation at sufficiently low temperatures, cold quantum gases are particularly intriguing.  

Whereas numerical approaches to thermalization can be based on a quantum Boltzmann collision term, it is interesting to have a simple, exactly solvable
model to better understand the physics of a given system. For bosons, such a model has been proposed in Ref.\,\cite{gw18}, and adapted to cold quantum
gases in Ref.\,\cite{gw18a}. The model is based on a nonlinear boson diffusion equation (NBDE) which has been derived from the Boltzmann collision term. For energy-independent
transport coefficients, it is structurally simple, but still complicated to solve exactly due to the nonlinearity in the drift term, which causes the system to reach the Bose-Einstein equilibrium distribution for sufficiently large times. 

Analytical solutions for physically meaningful nonlinear partial differential equations
are of great interest in several fields of physics, but are rarely available.
Notable examples in $1+ 1$ dimensions are the
Korteweg-de Vries equation \cite{bou72,kdv95}, which is of third
order in the spatial variable and has soliton solutions,
and Burgers' equation \cite{bur48}, which has the structure of
a one-dimensional Navier-Stokes equation without pressure term. It has been used to describe 
fluid flow and, in particular, shock waves in a viscous 
fluid, and it can be solved through Hopf's transformation \cite{ho50}. 
Although few
examples of completely integrable NPDEs in 2+1 dimensions are known -- such as the Kadomtsev-Petviashvili
\cite{kp70} and the Novikov-Veselov \cite{nv84} equations, which are
analogues of the KdV equation in two spatial dimensions
-- , higher-dimensional problems are usually not integrable.

In case of the nonlinear boson diffusion equation \cite{gw18,gw18a}, no exact
solutions in $2+1$ and higher dimensions are presently known.  
I discuss it here in $1+1$ dimensions (energy and time), which is appropriate for a cooling -- and eventually,  condensing -- isotropic three-dimensional thermal cloud of cold atoms. 
In Refs.\,\cite{gw18,gw18a} exact solutions were already obtained, but only for initial conditions confined to the energy domain $\epsilon\ge 0$, thus excluding the singularity
at $\epsilon = \mu < 0$. Without the singularity, however, the solutions do not reach the Bose-Einstein limit in the infrared for $t\rightarrow \infty.$

In order to attain a thermal equilibrium distribution both in the UV and IR energy regions, one therefore has to include the singularity in the initial conditions. It turns out that the chemical potential in the corresponding solutions is variabel, and can attain positive values.  To avoid such unphysical behaviour, one must specify boundary conditions at the singularity. This restricts the energy range in the initial conditions to energies larger than the chemical potential. Exact analytic solutions are still possible which have the proper Bose-Einstein equilibrium limit and, moreover, can represent evaporative cooling from an initial temperature $T_\text{i}$ to a final temperature $T_\text{f}$. 

With these exact solutions, I calculate the time-dependent entropy of an equilibrating finite Bose system and
compare to the results of a linear relaxation ansatz that enforces equilibration from the initial nonequilibrium distribution to a thermal distribution at a lower temperature {$T_\text{f}<T_\text{i}$.}

In the next section, the NBDE and its free analytic solutions \cite{gw18a} without and with consideration of the singularity are briefly reviewed. In the following section, the boundary conditions at the singularity are introduced, thus producing realistic physical results.
The derivation of the corresponding solutions is discussed in {Ref.\cite{rgw20}} since it exceeds the scope of a letter. Subsequently, evaporative cooling of cold bosonic atoms is modeled through the NBDE solutions.
With the exact solutions, the time-dependent entropy of a cooling Bose gas is finally calculated and discussed.

\section{Nonlinear diffusion equation and free solutions}
The transport equation 
for the single-particle occupation probability distributions $n\equiv n_\text{th}(\epsilon,t)$  has been derived from the bosonic Boltzmann collision term in Ref.\,\cite{gw18} as
 \begin{equation}
\frac{\partial n}{\partial t}=-\frac{\partial}{\partial\epsilon}\Bigl[v\,n\,(1+n)-n^2\frac{\partial D}{\partial \epsilon}\Bigr]+\frac{\partial^2}{\partial\epsilon^2}\bigl[D\,n\bigr]\,.
 \label{boseq}
\end{equation}
The drift term $v(\epsilon,t)$ accounts for dissipative effects, the term $D(\epsilon,t)$ for diffusion of particles in the energy space.
The many-body physics is contained in these
transport coefficients, which depend on energy, time and the second moment of the interaction.

{Spatial and momentum isotropy is a prerequisite for the reduction to $1+1$ dimensions, corresponding to \it{sufficient ergodicity}\rm\cite{lrw96}. For the thermal cloud of cold atoms around a Bose-Einstein condensate (BEC), this is expected to be a reasonable assumption, even though the condensate in a trap is spatially anisotropic. Concerning the role of different spatial dimensions in view of BEC formation, this enters my present formulation only through the density of states, which differs according to the number of spatial dimensions, and the confinement. The model calculations in this work are for a 3d system. One-dimensional systems where no BEC should be formed have not yet been investigated.}

In the limit of energy-independent transport coefficients the nonlinear boson diffusion equation  
for the occupation-number distribution $n(\epsilon,t)$
becomes
\begin{equation}
\frac{\partial n}{\partial t}=-v\,\frac{\partial}{\partial\epsilon}\Bigl[n\,(1+n)\Bigr]+D\,\frac{\partial^2n}{\partial\epsilon^2}\,.
 \label{bose}
\end{equation}
The thermal equilibrium distribution is a stationary solution 
\begin{equation}
n_\text{eq}(\epsilon)=\frac{1}{e^{(\epsilon-\mu)/T}-1}
 \label{Bose-Einstein}
\end{equation}
with the chemical potential $\mu<0$ in a finite boson system and $T=-D/v$. 
In spite of its simple structure, the NBDE with constant transport coefficients thus
preserves the essential features of Bose-Einstein
statistics which are contained in the bosonic Boltzmann equation. For a given initial condition $n_\text{i}(\epsilon)$, it can be solved exactly using the nonlinear transformation outlined
in Ref.\cite{gw18a}. The resulting solution can be written as
\begin{align}
    n(\epsilon,t) = -\frac{D}{v} \frac{\partial}{\partial\epsilon}\ln{\mathcal{Z}(\epsilon,t)} -\frac{1}{2}= -\frac{D}{v}\frac{1}{\mathcal{Z}} \frac{\partial\mathcal{Z}}{\partial\epsilon} -\frac{1}{2}
    \label{eq:Nformula} 
    \end{align}
 where the time-dependent partition function ${\mathcal{Z}(\epsilon,t)}$ obeys a linear diffusion equation
     \begin{align}
    \frac{\partial}{\partial t}{\mathcal{Z}}(\epsilon,t) = D \frac{\partial^2}{\partial\epsilon^2}{\mathcal{Z}}(\epsilon,t)\,.
    \label{eq:diffusionequation}
\end{align}
If no boundary conditions are specified, the free partition function becomes
    \begin{align}
    \mathcal{Z}_\text{free}(\epsilon,t)= a(t)\int_{-\infty}^{+\infty} G_\text{free}(\epsilon,x,t)\,F(x)\,\text{d}x\,.
    \label{eq:partitionfunctionZ}
    \end{align}
The physically more interesting case with boundary conditions  will be treated in the next section.
{The partition function} is only unique up to multiplication with energy-independent prefactors such as $a(t)$, since these drop out when taking the logarithmic derivative in the calculation of the occupation-number distribution.
The initial conditions that are contained in the function $F(x)$ cover the full energy region $-\infty<x<\infty$.

For a solution without boundary conditions as in Refs.\,\cite{gw18,gw18a}, Green's function \(G_\text{free}(\epsilon , x , t)\) of Eq.\,(\ref{eq:diffusionequation}) is a single Gaussian
\begin{align}
    G_\text{free}(\epsilon,x,t)=\exp\Bigl(- \frac{(\epsilon-x)^2}{4Dt}\Bigr)\,,
    \label{eq:Greensnonfixed}
\end{align}
but it becomes more complicated once boundary conditions are considered.
The function \(F(x)\) depends on the initial occupation-number distribution $n_\mathrm{i}$,
\begin{align}
    F(x) = \exp\Bigl[ -\frac{1}{2D}\bigl( v x+2v \int_0^x n_{\mathrm{i}}(y)\,\text{d}y \bigr) \Bigr]\,.
       \label{ini}
\end{align}
 The definite integral over the initial conditions taken at the lower limit drops out in the calculation of $n(\epsilon,t)$ when performing the logarithmic derivative. Hence, the integral can be replaced \cite{rgw20} by the indefinite integral $A_{\mathrm{i}}(x)$ over the initial distribution with $\partial_x A_{\mathrm{i}} (x) = n_{\mathrm{i}}(y)$, such that
\begin{align}
    F(x) = \exp\Bigl[-\frac{1}{2D}\left( v x+2v A_{\mathrm{i}}(x) \right)\Bigr]\,.
    \label{eq:G(x)}
\end{align}
It is now possible to compute the partition function and the overall solution for the occupation number distribution function Eq.\,(\ref{eq:Nformula}) analytically, even in the presence of a singularity in the initial conditions. (The singularity had been excluded in the initial conditions, and hence, in the solution given in Ref.\,\cite{gw18a}). 

As initial condition that is appropriate for a schematic description of evaporative cooling, one can start from a truncated thermal equilibrium distribution that is cut off at a maximum energy $\epsilon_\text{i}$ beyond which high-velocity atoms are removed 
\begin{equation}
n_\text{i}(\epsilon)=\frac{1}{e^{(\epsilon-\mu)/T}-1}\, \theta (1-\epsilon/\epsilon_\text{i})\,.
 \label{inibec}
\end{equation}
 If the integration in Eq.\,(\ref{eq:partitionfunctionZ}) is now carried out across the singularity at $x=\mu$, the solutions Eq.\,(\ref{eq:Nformula}) 
 approach a Bose-Einstein equilibrium distribution for $t\rightarrow \infty$. However, the chemical potential of the equilibrium solution moves to a larger value $\mu'>0$
\begin{equation}
\mu'=\frac{D}{v}\ln\Bigl[z^{-1}-\exp\,(-\epsilon_\text{i}/T)\Bigr]
 \label{mup}
\end{equation}
with the fugacity $z=\exp(\mu/T)$. Although this solution that includes the singularity is mathematically correct -- analytical and numerical results agree with high accuracy --, the shift has no reasonable physical interpretation, it is an artefact of the choice of the free Green's function Eq.\,(\ref{eq:Greensnonfixed}). Clearly, one has to consider the boundary conditions at the singularity in order to obtain physically meaningful solutions not only at short times when the step in the UV region is smeared out, but also in the IR at $t\rightarrow\infty$ when the thermal distribution is approached. These new solutions will be considered in the next section. Their detailed derivations are given in Ref.\,\cite{rgw20}.
\section{Exact solution with boundary conditions}
To solve the problem for constant temperature, but with boundary conditions at the singularity,  
the chemical potential is treated as a fixed parameter. With \(\lim_{\epsilon \downarrow \mu} n(\epsilon,t) = \infty\) \,$\forall$ \(t\), one obtains \( \mathcal{Z} (\mu,t) = 0\), and the energy range is restricted to  $\epsilon \ge \mu$. This requires a new Green's function 
that equals zero at \(\epsilon = \mu\) $\forall \,t$. It can be written as
\begin{align}
    {G} (\epsilon,x,t) = G_\text{free}(\epsilon - \mu,x,t) - G_\text{free}(\epsilon - \mu,-x,t)\,,
    \label{eq:newGreens}
\end{align}
and the partition function with this boundary condition becomes
\begin{align}
    {\mathcal{Z}} (\epsilon,t) = \int_0^\infty {G} (\epsilon, x, t)\,F(x+\mu)\, \text{d}x\,.
    \label{eq:newformulaforZ}
\end{align}
The function $F$ remains unaltered with respect to Eq.\,(\ref{eq:G(x)}), save for a shift of its argument by the chemical potential.

Using again Eq.\,(\ref{inibec}) as initial distribution, the occupation-number distribution 
can still be evaluated exactly in case of constant temperature, but also for $T_\text{i} \ne T_\text{f}$ as shown in the next section. The result for constant $T$ \cite{rgw20} 
\begin{align}
    {n}(\epsilon,t) = \frac{1}{\exp\Bigl(\frac{\epsilon-\mu}{T}\Bigr) L(\epsilon,t)-1}
    \label{eq:particledistributionfixedmu}
\end{align}
is formally similar to a Bose-Einstein distribution. The function \(L(\epsilon,t)\)  contains the terms that are responsible for the time evolution towards the equilibrium distribution at temperature $T=-D/v$. It can be written as
\begin{align}
    L(\epsilon,t)=\frac{\Sigma_1 (\epsilon ,t)}{\Sigma_2 (\epsilon ,t)}
\end{align}
with
\begin{align}
    \Sigma_1 (\epsilon,t)= \text{erfc}\Bigl({\frac{2\mu - \epsilon_i -\epsilon +t v}{\sqrt{4 D t}}}\Bigr)\notag - \exp\Bigl({\frac{\mu-\epsilon_i}{T}}\Bigr)\times \\ 
    \text{erfc}\Bigl( \frac{\epsilon_i-\epsilon +tv}{\sqrt{4 D t}}\Bigr)\,,
    \label{s1}
    \end{align}
    and
    \begin{align}
    \Sigma_2 (\epsilon,t) = \text{erfc}\Bigl({\frac{\epsilon-\epsilon_i +tv}{\sqrt{4 \, D \, t}}}\Bigr) \notag - \exp\Bigl({\frac{\mu-\epsilon_i}{T}}\Bigr)\times \\ 
    \text{erfc}\Bigl({\frac{\epsilon-2 \mu +\epsilon_i +tv}{\sqrt{4 D t}}}\Bigr)\,.
        \label{s2}
\end{align}
The complementary error functions are defined as
\begin{align}
    \text{erfc}\,(x) =1-\text{erf}\,(x) = \frac{2}{\sqrt{\pi}} \int_x^\infty e^{-t^2} \text{d}t\,.
\end{align}
As a consequence of the boundary condition, no unphysical shift in the chemical potential occurs in this special solution of the NBDE.
	 	\begin{figure}[t!]
	\centering
	\includegraphics[scale=0.435]{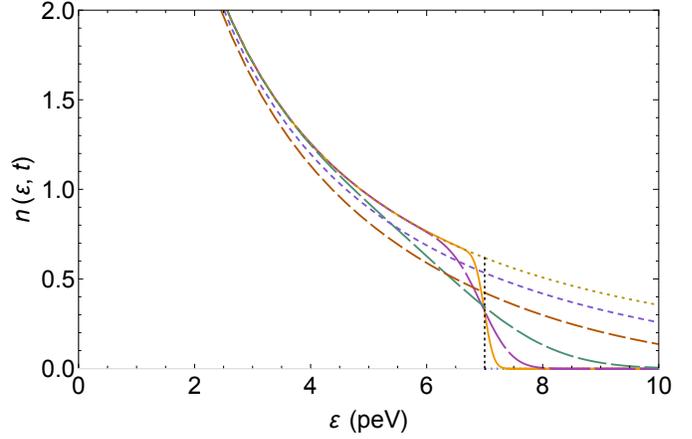}
	\caption{\label{fig1} (color online)  Equilibration of a finite Bose system based on the nonlinear evolution according to Eq.\,(\ref{bose})
	starting from a truncated Bose-Einstein distribution Eq.\,(\ref{inibec}), upper curve with cutoff at $\epsilon_\text{i}$\,=\,7\,peV$\simeq 1.7$ kHz.
	The transport coefficients are $D=8\times10^{3}$\,peV$^2$\,s$^{-1},~\,v=-1\times10^{3}$\,peV$\,$s$^{-1}$. The temperature $T=-D/v=8$\,peV\,$ \simeq93$\,nK
	is kept constant in this calculation.
The time evolution is shown at $t = 0.001, 0.01, 0.1, 1, 4$ and $40$\, ms (with decreasing dash length).
	}
	\end{figure}
The results for a specific parameter set that is adapted to cold quantum gases \cite{gw18a} are shown in Fig.\;\ref{fig1}.
To be able to compare directly with my earlier results in Ref.\,\cite{gw18a} that employed restricted initial conditions, the same parameters are used. The transport coefficients are $D = 8\times 10^3$ peV$^2\,$s$^{-1}$ and $v = -1\times 10^3$ peV\,s$^{-1}$, with an equilibrium temperature \(T = \SI{8}{\pico \electronvolt}\,\simeq 93\)\,nK. These values are motivated by experimental results for temperatures and time scales in ultracold $^{87}$Rb. At \(\epsilon_\text{i}= \SI{7}{\pico \electronvolt}\)$\,\simeq 1.7$ kHz, the initial thermal distribution is truncated, and the chemical potential is chosen as \(\mu = \SI{-0.68}{\pico \electronvolt}\). The temperature $T$ is kept constant in this particular calculation.
	\begin{figure}[t]
\centering
	\includegraphics[scale=0.435]{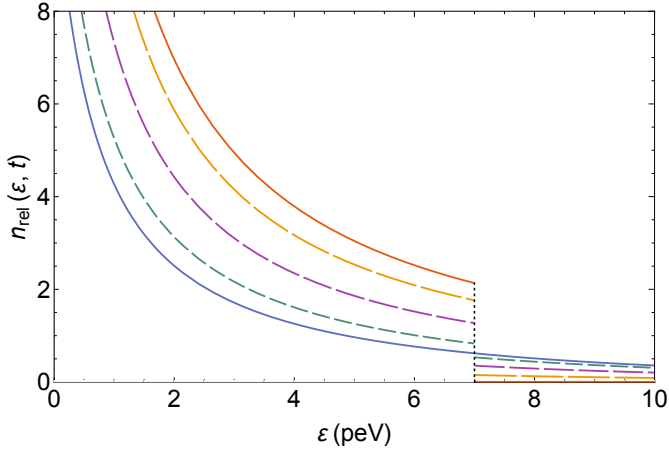}
	\caption{\label{fig2} (color online) Schematic representation of evaporative cooling in a bosonic system from temperature $T_\text{i}=20$ peV $\simeq 232$ nK (upper solid curve, truncated at 7 peV $\simeq 1.7$ kHz) to an equilibrium distribution with temperature $T_\text{f}=8$ peV $\simeq 93$ nK (lower solid curve). Discontinuous time-dependent single-particle occupation-number distribution functions at $t = 1, 3$ and $7$ ms (decreasing dash lenghts) are shown using the linear relaxation ansatz.
	}
	\end{figure}
		\begin{figure}[t]
	\centering
	\includegraphics[scale=0.435]{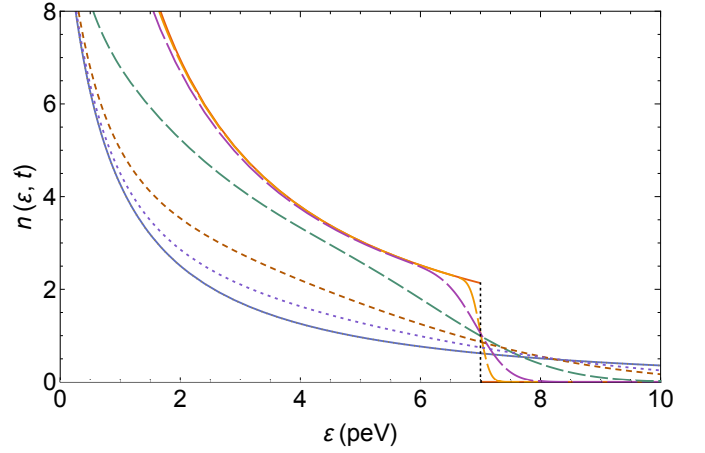}
	\caption{\label{fig3} (color online) Evaporative cooling in a bosonic system from temperature $T_\text{i}\simeq 232$ nK (upper solid curve, truncated at 1.7 kHz) to an equilibrium distribution with temperature $T_\text{f}\simeq 93$ nK (lower solid curve). Continuous time-dependent single-particle occupation-number distribution functions at $t = 0.001, 0.01, 0.1, 0.4,$ and $0.8$ ms (decreasing dash lenghts) are shown using exact solutions of the NBDE Eq.\,(\ref{bose}).
	}
	\end{figure}
	As is evident from the time-dependent analytical solutions shown in Fig.\;\ref{fig1} at $t = 0.001 - 40$ ms, the steep cutoff in the UV at $\epsilon = \epsilon_\text{i}$ is smeared out at short times as in the solution without boundary conditions. At larger times $t\gtrsim 1$ ms, the solutions do not approach a new equilibrium distribution with 
$\mu'>0>\mu$ as in the free case, but at $t\simeq 40$ ms return to the original equilibrium with $T=-D/v$, since this is the only temperature that exists in the present formulation. 

In the next section, different initial and final temperatures $T_\text{i} \ne T_\text{f}$ are considered to schematically account for evaporative cooling.
	\section{Analytical solutions for cooling}
	If the temperature $T_\text{i}$ in the initial conditions Eq.\;(\ref{inibec}) differs from the final equilibrium temperature $T_\text{f}=-D/v$ as is the case in evaporative cooling {\cite{an95,dmk95,lrw96}}, the analytic solutions of the NBDE with boundary conditions become more involved, but it is still possible to derive them. It is instructive to first consider the time evolution of the solutions from the initial nonequlibrium distribution $n_\text{i}(\epsilon)$ to the final equilibrium distribution $n_\text{eq}(\epsilon)$ in a linear relaxation ansatz
	$\partial\,n_\text{rel}/\partial t=(n_\text{eq} -n_\text{rel})/\tau_\text{eq}$ with 
\begin{equation}
n_\text{rel}(\epsilon,t)=n_\text{i}(\epsilon)\,e^{-t/\tau_\text{eq}}+n_\text{eq}(\epsilon)(1-e^{-t/\tau_\text{eq}})\,,
 \label{rela}
\end{equation}
which enforces equilibration towards the thermal distribution $n_\text{eq}(\epsilon)$
with the bosonic equilibration time $\tau_\text{eq}$, for which I use here the value $\tau_\text{eq}=4D/(9v^2)\simeq 3.6$ ms\,\footnote{This result refers to a $\theta$-function initial distribution,
see Ref.\cite{gw18}.}.
This simplified model can be used to compare with the 
nonlinear solution for cooling. 

Time-dependent results for the relaxation -time approximation (RTA) are shown in Fig.\,\ref{fig2}. 
Here, the initial distribution corresponds to a temperature of $T_\text{i}\simeq232$ nK, which is above the critical temperature for $^{87}$Rb.
The cut that accounts schematically for evaporative cooling is again placed at $\epsilon_\text{i}\simeq 1.7$ kHz. 
The equilibrium solution at a temperature $T_\text{f}=-D/v\simeq93$ K -- below $T_\text{crit}$ for reasonable density values -- 
is approached 
by the relaxation-time solutions that are shown at $t = 1, 3$ and $7$ ms. 

The new thermal distribution at the lower temperature is reached within $t\simeq 14$ ms. Due to the linear approximation of a nonlinear system, 
the RTA-solutions do not account for the correct physical behaviour of the system at the cut: The distribution functions remain discontinuous.

In contrast, the analytical solutions of the nonlinear NBDE produce a more realistic account of the thermalization. The partition function with boundary conditions at the singularity 
$\epsilon=\mu$ has been derived in Ref.\, \cite{rgw20} for $T_\text{i}\ne T_\text{f}$ as\\

\begin{align}
{\mathcal{Z}}(\epsilon,t) = \sqrt{4 D t} \, \exp\Bigl(-\frac{\mu}{2 T_{\mathrm{f}}}\Bigr) \sum_{k=0}^{\infty} \binom{\frac{T_{\mathrm{i}}}{T_{\mathrm{f}}}}{k} \left( -1 \right)^k \times  \notag \\
    \Bigg( \text{e}^{\alpha_k^2 D t} \left[ \text{e}^{\alpha_k (\epsilon - \mu)} \Lambda_1^k (\epsilon,t) - \text{e}^{\alpha_k (\mu - \epsilon)} \Lambda_2^k (\epsilon, t) \right] \notag \\
    + \exp\Bigl( \frac{(\mu - \epsilon_i)k}{T_{\mathrm{i}}}\Bigr) \exp\Bigl({\frac{D t}{4 T_{\mathrm{f}}^2}}\Bigr) \times \notag \\
  \Big[ \exp\Bigl(\frac{\epsilon-\mu}{2 T_{\mathrm{f}}}\Bigr) \Lambda_3 (\epsilon,t) - \exp\Bigl(\frac{\mu - \epsilon}{2 T_{\mathrm{f}}}\Bigr) \Lambda_4 (\epsilon,t) \Big] \Bigg)\,    \label{eq:Zarbtemp}
\end{align}
with $\alpha_k=1/T_\text{f}-k/T_\text{i}$, and the auxiliary functions
\begin{align}
    \Lambda_1^{k} (\epsilon,t) =& \,\text{erf}\,\Bigl(\frac{\epsilon - \mu +2 D t \alpha_k}{\sqrt{4 D t}}\Bigr) \notag  \\
        &\qquad -  \,\text{erf}\,\Bigl(\frac{\epsilon - \epsilon_i + 2 D t \alpha_k}{\sqrt{4 D t}}\Bigr)\,,\\
    \Lambda_2^{k} (\epsilon,t) =& \,\text{erf}\,\Bigl(\frac{\mu-\epsilon+ 2 D t \alpha_k}{\sqrt{4 D t}}\Bigr) \notag \\ 
    &\qquad -  \,\text{erf}\,\Bigl(\frac{2 \mu - \epsilon - \epsilon_i + 2 D t \alpha_k}{\sqrt{4 D t}}\Bigr)\,,\notag\\
\end{align} 
\begin{align}   
    \Lambda_3 ( \epsilon,t) =& \,\text{erfc}\,\Bigl(\frac{\epsilon_i - \epsilon + t v }{\sqrt{4 D t}}\Bigr)\,, \\
    \Lambda_4 ( \epsilon,t) =& \,\text{erfc}\,\Bigl(\frac{\epsilon - 2 \mu + \epsilon_i + t v}{\sqrt{4 D t}}\Bigr)\,.
\end{align}
The derivative $\partial  {\mathcal{Z}}/\partial\epsilon$ can also be obtained analytically, such that the time-dependent occupation-number distribution function for evaporative cooling can be calculated from Eq.\,(\ref{eq:Nformula}). For $T_\text{i}=T_\text{f}$, the result of Eq.\,(\ref{eq:particledistributionfixedmu}) is recovered.

The time-dependent analytic distribution functions that solve the NBDE exactly with boundary conditions at the singularity are displayed in Fig.\;\ref{fig3} for the same initial conditions as in Fig.\;\ref{fig2}. They agree precisely with numerical solutions of the basic equation, {and predict the time-dependent cooling from a thermal distribution with temperature $T_\text{i}$ that is truncated at $\epsilon_\text{i}$ to a BE distribution with $T_\text{f} < T_\text{i}$, which is the thermal distribution for $t\rightarrow \infty$. This is similar to the kinetic theory of evaporative cooling in works such as Ref.\,\cite{lrw96}, but now an analytic aolution is given.}

{The values of the transport coefficients $v, D$ in this specific model calculation have been derived from their relations to the equilibrium temperature $T=-D/v$ and the equilibration time $\tau_\text{eq}=4D/(9v^2)$ \cite{gw18}, with $T=93$ nK and $\tau_\text{eq}=3.6$ ms. For future direct comparisons with experiment, these values shall be adapted to the corresponding data.}

Thermalization in the infrared occurs faster than in case of the linear relaxation ansatz: For the present parameter set, the thermal distribution is reached within $t\simeq 1$ ms in the IR.  The buildup of the thermal slope in the UV is, however, slower in the nonlinear model as compared to the relaxation ansatz, which enforces {a rapid} approach to {the} Boltzmann-like tail. 

{It would be instructive to plot also the rate of increase of atoms in the condensate $N_\text{c}(t)$, for conserved total particle number $N = N_\text{c}(t)+N_\text{th}(t)$, and $T<T_\text{c}$. This requires, however, to go beyond the above exact analytic solutions, because these are derived for constant chemical potential, whereas particle-number conservation -- which is a necessary condition for condensate formation to occur -- necessitates a time-dependent chemical potential. With $\mu\rightarrow\mu(t$), new nonvanishing terms arise when taking the time derivative, such that the exact analytical solutions of the NBDE are only approximately valid. Moreover, a self-consistent approach needs a quantum treatment of $N_\text{c}(t)$, which is beyond the scope of the present nonequilibrium-statistical work.
 We have discussed this in more detail in Ref.\,\cite{rgw20}, with a numerical determination of $\mu(t)$}.

\section{Time-dependent entropy}
A condition that physically reasonable solutions of the nonlinear boson diffusion equation should fulfil is the increase of the entropy with time towards the equilibrium value that is determined by the final Bose-Einstein distribution. This is indeed the case for the solutions of the NBDE at constant temperature $T$ shown in Fig.\;\ref{fig1}: The entropy $S_\text{i}$ of the initial nonequilibrium distribution rises monotonically in the course of the time evolution to the final equilibrium entropy $S_\text{eq}$ by about a factor of three. 

In a cooling system as displayed in Figs.\;\ref{fig2} and \ref{fig3}, however, the time dependence of the entropy $S(t)$ is more involved due to the interplay of cooling, which tends to decrease the entropy, and thermalization, which causes an enhancement. In the example shown there with $T_\text{i}\simeq 232$ nK and $T_\text{f}\simeq 93$ nK, the entropy of the initial thermal distribution -- without the cut -- is about four times larger than the one of the final distribution, whereas the initial nonequilibrium distribution -- with the cut at $\epsilon=\epsilon_\text{i}$ -- carries only half the entropy of the final equilibrium distribution at the lower temperature $T_\text{f}$. The detailed time dependence of $S(t)$ is then subject of a corresponding model calculation.

If the final temperature  $T_\text{f}$ is below the critical value for condensate formation $T_\text{crit}$ such that particles occupy the condensed state, the total entropy still equals the entropy of the atoms in the thermal cloud. As emphasized in Ref.\,\cite{scul18}, this is the case even though the entropy of the particles in the ground state is nonzero, because the latter is cancelled by the so-called correlation entropy due to the fixed number of particles distributed among the quantum states. 

It is therefore sufficient for a calculation of the total time-dependent entropy $S(t)$ to consider only the thermal cloud. For the corresponding numerical calculation, the analytical solutions based on Eqs.\,(\ref{eq:Nformula}) and (\ref{eq:Zarbtemp}) for constant chemical potential $\mu$ and boundary conditions at the singularity $\epsilon=\mu$  are used. The  entropy in a bosonic system
for an average number of particles $n(\epsilon,t)$ per single-particle state can be written as
 \cite{yam86}
\begin{eqnarray}
S(t)=\int_0^\infty g(\epsilon)\Bigl[\ln\bigl(1+n(\epsilon,t)\bigr)\qquad\qquad\notag\\
+n(\epsilon,t)\ln\bigl(1+1/n(\epsilon,t)\bigr)\Bigr]\text{d}\epsilon\,,
\label{entropy}
\end{eqnarray}
where the density of states for a three-dimensional isotropic Bose gas without external potential obeys the power law
\begin{align}
    g(\epsilon) = g_0 \, \sqrt{\epsilon}
\end{align}
with $(\hbar=c=k_\text{B}=1)$
\begin{align}
g_0=(2m)^{3/2}\,V/(4\pi^2)\,,
\end{align}
as obtained from the substitution of a summation over the quantum numbers of the associated states with an energy integration \cite{BookPitaevskii}.

{The spatial dimensionality and the external confinement thus enter the present formulation only through the density of states, enabling future considerations of their effect 
on BEC formation. In particular, one could try to verify that condensate formation does not occur in a 1d box because of the scaling of the density of states, but this requires to consider the solutions for time-dependent chemical potential, as well as the properties of the trapping potential as discussed in Ref.\,\cite{dmk95}}.

The first term {in the entropy Eq.\,(\ref{entropy})}  is usually referred to as \it{wave entropy}\rm, it yields the largest contribution when the single-particle state is occupied by many particles, as in the IR. The second term is the \it{particle entropy}, \rm  which is more relevant in case of low occupation $n(\epsilon,t)<1$, as in the UV. 

\begin{figure}[t]
	\centering
	\includegraphics[scale=0.435]{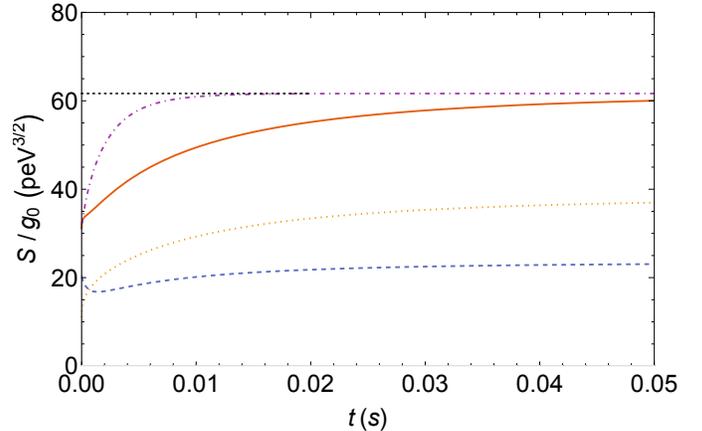}
	\caption{\label{fig4} (color online) Time evolution of the entropy $S(t)/g_0$ in an equilibrating Bose system in the course of evaporative cooling from $T_\text{i}=232$ nK to $T_\text{f}=93$ nK as calculated
	from the analytical solution of the NBDE Eq.\,(\ref{bose}) with constant chemical potential $\mu$, solid curve. {$S(t=0)$ is the total entropy following evaporative cooling}. The dashed curve is the wave entropy, the dotted curve the particle entropy, and the dot-dashed curve the result from the linear relaxation ansatz. The dotted horizontal line indicates the equilibrium value {at $T_\text{f}=93$ nK}. The entropy of the analytical solution reaches the thermal value at $\simeq 60$ ms.}
	\end{figure}

Results for the time-dependent entropy are shown in Fig.\,\ref{fig4}, with the same parameters as in Fig.\,\ref{fig3} for the analytical solutions of the NBDE. Here, the equilibrium value of the entropy at the initial temperature $T_\text{i}\simeq 232$ nK is $S_\text{eq}/g_0\,(T_\text{i})\simeq256.35$ peV$^{3/2}$. With a sharp cutoff at $\epsilon_\text{i}=7$ peV $\simeq 1.7$ kHz
to account for evaporative cooling, the entropy of the initial (cooled) nonequilibrium distribution is reduced to  $S_\text{i}/g_0\,(T_\text{i})\simeq30.96$ peV$^{3/2}$. 

Thermalization during the time evolution then occurs through the analytical solutions of the NBDE, and the result for the rising entropy $S(t)/g_0$ is shown in the solid curve
in Fig.\,\ref{fig4}. 
The new equilibrium value of the entropy at the final temperature $T_\text{f}\simeq 93$ nK following evaporative cooling and thermalization is $S_\text{eq}/g_0\,(T_\text{f})\simeq61.66$ peV$^{3/2}$, dotted horizontal line.

The dashed curve is the wave entropy. It is most relevant for large occupation numbers, which are present at all times in the IR region $\epsilon<\epsilon_\text{i}$, and therefore, this contribution shows a rather weak time dependence. The dotted curve is the particle entropy, which is initially smaller than the wave entropy, because the occupation in the UV beyond the cut
is negligible at small times. In the course of thermalization, however, it quickly exceeds the wave entropy at $t\simeq 0.7$ ms for the parameter set used in this work, and rises subsequently. Hence, the entropy at large times is mostly determined by the contribution of the thermal tail.

Whereas the analytical solutions of the NBDE reach the thermal equilibrium values in the IR rather fast within about 1.4 ms as shown in Fig.\,\ref{fig3}, it takes more time to 
build up the thermal tail in the UV beyond the cut: The rise of the particle entropy in the nonlinear model occurs fairly slowly, such that the total entropy reaches the equilibrium value only within about 60 ms.

These results for the time-dependent entropy as obtained from the analytic solutions of the NBDE are compared in Fig.\,\ref{fig4} with the corresponding relaxation ansatz, dot-dashed curve. 
Although thermalization in the IR occurs more slowly when using this ansatz as compared to the nonlinear model -- see Figs.\,\ref{fig2} and \ref{fig3} --, the exponential Boltzmann-like slope in the UV appears faster.
The rapid rise of the entropy at relatively short times in the relaxation-time approximation is a consequence of this enforced fast buildup of a thermal tail in such a linear approach. 
As expected, the system's entropy approaches the same equilibrium value, but at shorter times than the nonlinear solution.

\section{Conclusion}
New exact solutions of the nonlinear boson diffusion equation have been explored which take account of the singularity in the initial conditions at $\epsilon=\mu<0$, and the necessary boundary conditions at the singularity. Different from earlier results that were calculated with the free Green's function, these solutions converge towards the Bose-Einstein equilibrium.
 {Hence, they properly account for thermalization not only in the UV, but also in the IR region}. The {analytic} solutions are in excellent agreement with numerical approaches.

The analytic results of the NBDE are tailored to describe equilibration processes that occur in quantum gases in the course of evaporative cooling, and subsequent re-thermalization.
They are applied schematically to bosonic atoms like $^{87}$Rb and their evaporative cooling, which is a precondition for condensate formation. For cooling from an initial temperature $T_\text{i}$ to a final temperature $T_\text{f}$, the average single-particle occupation-number distributions are calculated as function of time, and compared to a relaxation ansatz.

Using both the linear and the nonlinear model of thermalization, the time-dependent entropy is calculated, and the contributions of wave and particle entropy are discussed.
Whereas cooling drastically reduces the entropy, the subsequent re-thermalization causes a gradual increase of the entropy towards the equilibrium value, which coincides with the Bose-Einstein result and is significantly below the initial value before cooling. 

Further refinements of the model such as variable transport coefficients {and time dependent chemical potentials} are conceivable, but may not allow for analytic solutions. Extensions of the NBDE itself to higher dimensions in order to account for anisotropic systems should also be investigated. Direct comparisons of the results to data from cold-atom experiments would be most welcome.

\acknowledgments
Discussions with Johannes H\"olck about NBDE-solutions with boundary conditions,  Niklas Rasch in the course of his BSc thesis and Ref.\,\cite{rgw20}, 
and Alessandro Simon regarding a detailed comparison of the analytic solutions with numerical results are gratefully acknowledged.
\bibliographystyle{eplbib}
\bibliography{gw_20}

\end{document}